\documentclass[letterpaper,twocolumn,showpacs,preprintnumbers,amsmath,amssymb]{revtex4-1}

\usepackage{graphicx}
\usepackage{dcolumn}
\usepackage{bm}

\DeclareGraphicsExtensions{.eps,.EPS,.pdf}

\def\unit#1{\mathop{\rm #1}\nolimits}%

\begin{document}


\title{Fabrication of an atom chip for Rydberg atom-metal surface
interaction studies}

\author{O.~Cherry} \altaffiliation{Deceased.}
\author{J.~D.~Carter}
\author{J.~D.~D.~Martin}
\affiliation{
Department of Physics and Astronomy and Institute for Quantum Computing,
University of Waterloo, Waterloo, Ontario, N2L 3G1, Canada
}

\date{\today}

\begin{abstract}
An atom chip has been fabricated for the study of interactions between $^{87}$Rb Rydberg atoms and a Au surface. The chip tightly confines cold atoms by generating high magnetic field gradients using microfabricated
current-carrying wires. These trapped atoms may be excited to Rydberg states at well-defined atom-surface distances. For the purpose of Rydberg atom-surface interaction studies, the chip has a thermally evaporated Au surface layer,
separated from the underlying trapping wires by a planarizing polyimide dielectric. Special attention was paid to the edge roughness of the trapping wires, the planarization of the polyimide, and the grain structure of the Au surface.
\end{abstract}



\maketitle

\textbf{Note:} This is a manuscript prepared approximately 13 years ago, intended to be combined with experimental demonstration of the chip to trap atoms.  The first author passed away in September 2024.  The fabrication techniques may be useful to others.

\section{Introduction\label{intro}}

A Rydberg atom has a single electron excited to a state of high
principal quantum number $n$. Rydberg atoms are remarkable -- compared
to less excited atoms -- for their exaggerated properties,
stemming from a large average separation of their excited
electron and ion-core \cite{gallagher:1994}.  In particular, they
are highly susceptible to electric fields: their polarizabilities
scale like $n^7$, and the field required for removal of the excited electron
(ionization) scales like $1/n^4$.

Rydberg atoms also exhibit enhanced interactions with metal surfaces.
The image charge picture is useful for understanding
these interactions \cite{chaplik:1968}.
Energy shifts due to the interaction of the Rydberg atom with its
image dipole scale like $n^4/z^3$, where $z$ is the distance to
the surface.  This has been observed spectroscopically by
Sandoghdar {\it et al.}~\cite{sandoghdar:1992}.  Closer to
the surface, the interaction becomes highly non-perturbative,
and the Rydberg atom may be field ionized by its own image.
Hill {\it et al.}~\cite{hill:2000} have
experimentally verified that this occurs at an atom-surface
distance of $4.5 a_0 n^2$, where $a_0$ is the Bohr radius.

There have been two notable challenges in
previous Rydberg atom surfaces experiments:
1) control and/or determination
of the atom-surface distance $z$, and 2) minimization of
the influence of stray surface electric fields.
This paper describes the fabrication of an ``atom chip''
\cite{folman:2002,fortagh:2007}
suitable for Rydberg atom-surface interaction studies
at {\em controllable} separations.  We discuss its use for
characterizing stray electric fields close to surfaces.

Atom chips are based on the principle
that paramagnetic atoms
experience mechanical forces in inhomogeneous magnetic fields.
In particular, atoms in low-field seeking states are
drawn to regions of low magnetic field magnitude.
Micron-scale current-carrying wires lithographically
patterned on a substrate
(known as ``magnetic microtraps'' or ``atom chips'')
can be used to produce local magnetic
field minima
\cite{folman:2002,fortagh:2007}).
These chips can trap
translationally cold atoms obtained from laser and
evaporative cooling at variable distances from their surfaces.
The distance from the surface
may be varied by changing the currents in the wires generating
the potentials.
For example, atom chips have been used to observe
the Casimir-Polder force between ground-state atoms and
dielectric surfaces -- this force reduces
the well-depth for the trapped atoms, causing increased atom loss
as the surface is approached \cite{lin:2004}.

Initial work on Rydberg excitation using atom chips near surfaces
has shown large inhomogeneous fields due to adsorbed Rb
\cite{PhysRevA.81.063411}.  However, even without adsorbed contaminants we expect to observe inhomogeneous patch fields due to the polycrystalline nature of the metal surface.  We plan to measure these ``patch fields'' as a function of distance from the surface by exciting magnetically microtrapped atoms.
The spectral broadening of these transitions will indicate the
patch field magnitudes.  We have designed and fabricated an
atom chip for this purpose.

\section{Design}

In this section we discuss the design rationale for the atom-chip.
Use of the chip for Rydberg atom-surface
experiments requires: a source of cold atoms, the capability
to load them into the trap, and equipment for the excitation
and detection of Rydberg atoms.  These aspects of the experiment are
discussed in Appendix \ref{appendix:experimentalsystem}.

\begin{figure}
\includegraphics{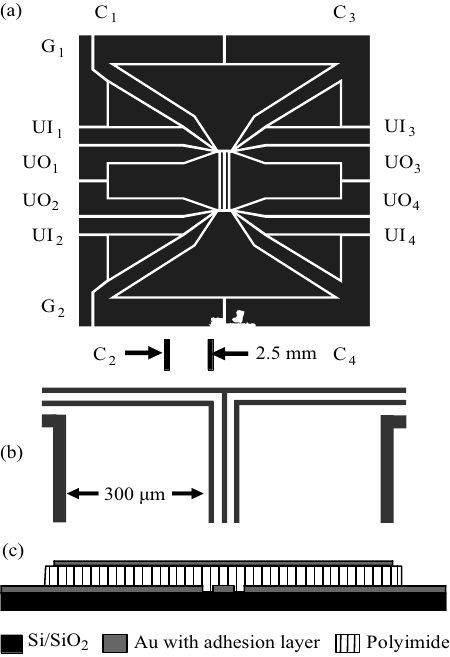}
\caption{\label{f1}
Schematic of the atom chip. (a) Trapping wires (b) Magnified view
of the end of the center wire strip is shown at
right. The widths of the center wires and spaces
are $7 \unit{\mu m}$, but
are exaggerated here for clarity. (c) Cross section of the chip.
}
\end{figure}

There are five parallel trapping wires at the center of
the chip (see Fig.~\ref{f1}a).  These are approximately $4\unit{mm}$
long.  The center wire is $7 \unit{\mu m}$  wide, sufficiently narrow to form
a tight trap close to the wire.\ It is an H-shaped wire, allowing it
to function as either a Z- or U-wire (C$_{1-4}$). A $7 \unit{\mu m}$
wide
U-shaped wire sits on either side of the center wire, separated
from the center wire by a $7 \unit{\mu m}$ gap (UI$_{1-2}$, UI$_{3-4}$).
A $14 \unit{\mu m}$ wide U-wire neighbors each of the inner U-wires
(UO$_{1-2}$, UO$_{3-4}$), separated by $300 \unit{\mu m}$
(see Fig.~\ref{f1}b).
Each wires terminates in a bond pad at the edge of the chip.

To load the chip, will form a trap
far away from the chip surface ($z > 200 \unit{\mu m}$),
by running currents in the same direction in the center wire and inner
U-wire, and flowing in the opposite direction in the outer U-wire.
For traps at intermediate distances ($50 < z < 200 \unit{\mu m}$)
the center wire and outer U-wire will have currents flowing
in opposite directions (and the inner U-wire not used).
Close to the chip ($z < 50 \unit{\mu m}$) traps will be
created with the
center wire current opposing the current in the inner U-wire
(and the outer U-wire not used).

The atom chip allows precise control of the atom surface distance.
However, there is some uncertainty in this distance due to the
finite size of the trapped atom cloud.  This is detailed in
Appendix \ref{appendix:control}.

The current-carrying wires responsible for magnetic trapping will
have appreciable voltage drops along their lengths and voltage differences
between them.  If unshielded,
these would create strong electric field
inhomogeneities over the trapped atom sample.  As Rydberg atoms are
especially sensitive to electric fields, this is undesirable.
In addition, exposed dielectric between the wires may charge up,
creating time dependent electric fields.  To avoid these problems
it is necessary to have a ``shield layer'' between the wires and the
atoms.
To prevent charging, this shield should be
conductive (a metal or doped semiconductor), but insulated
from the trapping wires below.
For mechanical rigidity and
durability it makes sense to use an insulating dielectric layer.
However, this layer should not be too thick, as it will increase
the distance between the trapping wires and atoms, reducing the
trap confinement.
The trap should be as confining as possible,
so that the atom-surface distance is well-defined.

It is also desirable for the insulating layer to ``planarize''
the underlying wire topography.  The shield layer
should be as flat as possible, presenting an idealized surface to the
atoms, which will facilitate comparison with theory.  For this reason
we will refer to the intermediate, insulating layer between the wires
and shield as the ``planarization layer''.

A flat, reflective surface is often required above an atom chip
to form a mirror MOT \cite{reichel:1999}.  For example,
Reichel {\it et al.}~\cite{reichel:2001} have used a replica
transfer technique.  A chip was covered in epoxy, and then
a flat substrate with a thin silver layer (a mirror) was placed face
down on the chip, separated from its surface by spacers.
Once the epoxy cured, the substrate was lifted off, leaving
a flat silver layer on top of the chip.   The epoxy layer was
approximately $25 \unit{\mu m}$ thick -- which significantly reduces the
confinement possible near the chip surface.  However, it is possible that
modifications of this approach could produce thinner planarization
layers.

An improvement to these methods is to coat the wires with a planarizing
dielectric polymer. Both polyimide and benzo-cyclo-butene (BCB) have
been used to obtain flatter surfaces
\cite{thesis:lev}, \cite{treutlein:2006}, \cite%
{article:harochesuperconductingchip}, \cite{article:estevecorrugation2}.
Their application and patterning are compatible with standard clean room
practices and fabrication equipment. In this work we chose to use
polyimide PI 2562
from HD Microsystems, with a single-layer thickness of $1.4 \unit{\mu m}$
when spun at $3000 \unit{RPM}$.
Additional layers can be deposited either before or after curing the film.

The planarization requirement may be lessened by submerging the
trap wires in the substrate \cite{treutlein:2006}.
This has the additional advantage
of improving thermal conduction from the wires.  A possible
limitation of this technique is the edge roughness of the wires.
Despite the additional processing required, this approach is worth
further investigation.

As Rydberg atoms will directly interact with the shield, its properties
are important.  Gold was chosen for its chemical inertness and ease
of thermal evaporation.  As was mentioned in the Introduction, a
major challenge in Rydberg atom experiments are the
stray electric fields produced by the surface.  These may
have several origins.  As discussed in Appendix \ref{appendix:patch},
so-called
``patch fields'' are produced by polycrystalline surfaces typical
of thermal evaporation of thin gold films.  To minimize these patch
fields it is desirable to minimize grain size.  This can be influenced
to some degree by the evaporation conditions and film thickness, as
we shall discuss later.

\section{Fabrication\label{design_fab}}

The substrate used in our atom chip fabrication is Si with a $40 \unit{nm}$
thermally-grown SiO$_{2}$, which insulates electrically, and has
been shown to be thin enough to provide sufficient thermal contact
to the substrate \cite{groth:2004}. Before
patterning, wafers are cleaned using the RCA 1 process
(5:1:1 H$_{2}$O:NH$_{4}$OH:H$_{2}$O$_{2}$ \cite{kern:1970}) to
remove unwanted organic contamination that may have occurred during
storage.

The dimensions of the chip are $2.02\times 2.02\unit{cm}$ --
constrained to the largest size that will fit through a
2.75 in Conflat port on the vacuum chamber.

The wires are patterned using AZ 2035 $n$LOF negative photoresist
spun at $2000 \unit{RPM}$ to give a thickness after baking of between 3.5
and $4.0 \unit{\mu m}$.\ The resist process follows the recommendations
of the manufacturer \cite{datasheet:AZ_2000} except developing
time is increased to $120 \unit{s}$ to increase undercut into the sidewalls.

Metal films (for both the wires and surface shield)
are deposited using an Edwards E306A thermal deposition
system. The system uses a diffusion pump to reach a base pressure
of between 7 $\times$ 10$^{-7}$ and 3 $\times$ $10^{-6} \unit{Torr}$.
Film thickness is monitored using a $6 \unit{MHz}$ quartz crystal oscillator.
For reasons discussed later, we use a Ti/Pd/Au adhesion layer for
the wires, and a Cr/Au adhesion layer for the shield.
We evaporate Cr from a tungsten rod electroplated with a thick layer
of Cr. Ti is evaporated from a cylindrical pellet sitting in a
tungsten boat. Pd is evaporated from cylindrical pellets with a
tungsten coil and alumina crucible. The source of Au is
a $1 \unit{Oz}$ bar cut into small pieces. Au is
evaporated in an alumina crucible heated by a tungsten coil.
In order to minimize radiative heating which may damage the
resist, the substrate mount is water cooled.

The corresponding remover for AZ 2035 $n$LOF resist is AZ Kwik Strip.
However, the first stage of lift-off is in warm acetone. Hot Kwik Strip
is then used to remove baked-on resist residue. A low power
ultrasonic clean in isopropanol removes metal flakes that may
have clung to the surface.

The chip is planarized using PI 2562 polyimide.
An adhesion promoter, VM 652, is first spun over the chip at
$3000 \unit{RPM}$
and baked on a hotplate at $120 \unit{^\circ C}$ for $30 \unit{s}$.
A $2 \unit{cm}$ disk of
PI 2562 is poured onto the chip and spun at $500 \unit{RPM}$
for $10 \unit{s}$
followed by $3000 \unit{RPM}$ for $30 \unit{s}$.
It is baked immediately after
spinning at $120 \unit{^\circ C}$ for $5 \unit{min}$.
The polyimide is cured in a nitrogen
fed furnace at $200 \unit{^\circ C}$ for $30 \unit{min}$
and $350 \unit{^\circ C}$ for
$60 \unit{min}$
with ramps of
$4 \unit{^\circ C/min}$ and
$2.5 \unit{^\circ C/min}$ ramps,
respectively. A single layer of cured PI 2562 has a thickness of
$1.3-1.5 \unit{\mu m}$. To increase planarization, the chip is coated
with two additional layers of polyimide, with a full cure between each.
Multiple layers of PI 2562 may be coated before curing, but this has
been found to result in a lower degree of planarization.

\begin{figure}
\includegraphics{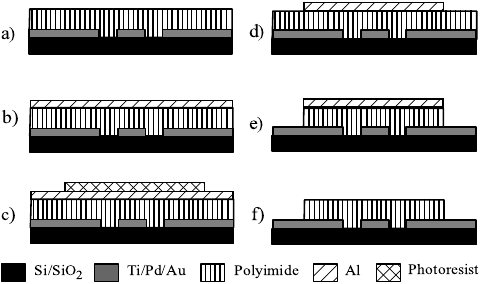}
\caption{\label{f2}
Patterning of polyimide planarization layer: a) coat and cure polyimide;
b) sputter Al; c) pattern positive photoresist over Al;
d) wet etch Al using photoresist as mask, then remove photoresist;
e) reactive ion etch (RIE) polyimide using Al as mask;
f) wet etch to remove remove Al mask.
}
\end{figure}

The polyimide must be
removed around the perimeter of the chip
to reveal the bond pads of the underlying wires.
The polyimide is patterned using ICP-RIE (Trion Phantom II) with
O$_2$ as the etch gas. The patterning process is shown in
Fig.~\ref{f2}. A sputtered Al layer, $0.5-1 \unit{\mu m}$
thick, acts as the
etch mask. It is patterned using AZ 3312 positive photoresist and
etched in PAN etch
($1$:$16$:$1$:$2$ H$_{3}$PO$_{4}$:CH$_{3}$COOH:HNO$_{3}$:HNO$_{3}$:H$_{2}$O)
at $40-45 \unit{^\circ C}$.
The polyimide is etched in ICP/RIE at a pressure of
$10 \unit{mTorr}$ and an O$_2$ flow rate of $35 \unit{sccm}$.
An ICP power of $500 \unit{W}$
and a RIE power of $25 \unit{W}$ gives an etch
rate of about $0.85 \unit{\mu m/min}$.
After the etch, the Al is removed in PAN.

In order to avoid etching unwanted pinholes through the polyimide,
coating and etching is performed twice. The chip is first coated
with two layers of polyimide (with a cure after each coating) which
are then etched. The chip is then coated with a third layer of
polyimide which is patterned using the same process as the first two.
Since the depth of this second etch is less than that of the first,
a hole cannot be etched through all three layers.

The shield layer is deposited using thermal evaporation and patterned
with lift-off lithography using the same process as the wires. The
Au thickness is $100 \unit{nm}$ with a $12-20 \unit{nm}$
Cr adhesion layer. The shield
is connected to ground pads G$_{1}$ and G$_{2}$
(see Fig.~\ref{f1}a) using small drops of Epotek H21D silver-filled epoxy.

The chip is mounted on a machined Macor block using Epotek 353ND epoxy.
Electrical connections between the chip and Au pads on the mount
are made using ultrasonic wedge bonding with
$25 \times 250 \unit{\mu m}$
Au ribbon. Each connection uses two ribbons to reduce the current
passing through a ribbon, and to add security in case a ribbon breaks
or a bond becomes unstuck. The strength of each bond is enhanced with
a bead of Epotek H21D epoxy.

\section{Results\label{results}}

Several measurements and tests of the chip have been made to ensure that
it will perform satisfactorily. Wire heating has been measured to be very
similar to the measured values of chips fabricated by Groth \textit{et al.}
\cite{groth:2004} both at fast timescales (dependent on the contact
resistance
between the wires and substrate) and slow timescales
(dependent on  the thermal
conductivity of the substrate). In air, the chip can support current densities
of above $9 \times 10^{6} \unit{A/cm^{2}}$
with a $500 \unit{ms}$ pulse length and is
expected to perform similarly under vacuum.

\begin{figure}
\includegraphics{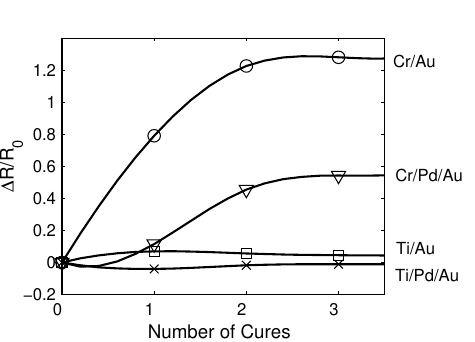}
\caption{\label{f3}
Change in resistance in the center wire with Cr/Au, Ti/Au,
Cr/Pd/Au and Ti/Pd/Au
metallizations after subjecting to polyimide cure cycles.
The solid lines are to aid the eye.
}
\end{figure}

As was mentioned in the previous section we use Ti as an
adhesion layer with a Pd
intermediate diffusion barrier for the Au wires.
We have measured interdiffusion effects for
various metallizations. The
trapping wires of early chips were fabricated using Cr as the adhesion
layer. It was observed that, after the subsequent high temperature
processing of the polyimide layer, the resistance in the wires
increased by over 120\% due to interdiffusion between the Cr and Au.
This effect was very similar to that measured by Madakson
\cite{article:madakson_diffusion} and Munitz \textit{et al}$\mathit{.}$
\cite{Munitz:diffusion} who showed that resistance increases as the Cr
diffuses into the gold film through grain boundaries at temperatures
above $300 \unit{^\circ C}$. Once it reaches the surface it forms an oxide,
Cr$_{2}$O$_{3}$. The resistance reaches a maximum as Cr saturates
the Au film and then drops off once the Cr layer becomes depleted.
Eventually the majority of Cr is converted to Cr$_{2}$O$_{3}$ and the
resistivity decreases.\ The measurement of this chip indicated that,
after three polyimide cure processes, resistance had reached the
saturation point,
and that the Cr adhesion layer was not yet depleted. As a test,
chips were also fabricated
using Ti/Au and Cr/Pd/Au metallizations. Figure \ref{f3} plots the
increase in wire resistance after successive cure cycles for each
metallization. As it can be seen the Cr/Au and Cr/Pd/Au metallizations
exhibited the poorest performance with resistance increasing by over
125\% and 55\% respectively. The resistance of the Ti/Au wires increased by
0.5\% and the Ti/Pd/Au wire resistance \textit{decreased} by approximately 1\%.

Another concern with diffusion is the buildup of oxide on the wire
surface.  This causes problems when making electrical connections by
wirebonding.\ The ability to make reliable, adherent bonds on Cr/Au
bond pads with gold ribbon was found to decrease significantly
after a $400 \unit{^\circ C}$ bake by
Pan \textit{et al}$\mathit{.}$ \cite{conference:Pan_diffusion}.
Our attempts to wirebond to the Cr/Au contacts after curing meet
with the same problems. The wires can be easily pulled from the
surface even if they appear to stick after making the bond.
In contrast, bondability to Ti/Pd/Au chips is excellent.

\begin{figure}
\includegraphics{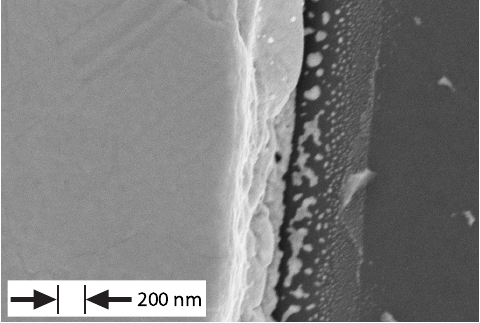}
\caption{\label{f4}
SEM image of a wire edge.
To the right of the wire is a thin layer of
metal that reached the wafer at a non-normal incident
angle and
passed beneath
the undercut of the photoresist.
This film exists exhibits the characteristics of island
formation in preliminary thin film growth \cite{Liu:1997}.
}
\end{figure}

The edge roughness of a typical wire is shown in Fig.~\ref{f4} .
This chip has been submitted
to three cures at $350 \unit{^\circ C}$, which has had an annealing effect
on the metal. Before curing, the wire edge roughness is limited to
deviations of about $100 \unit{nm}$ by the edges of the photoresist.
After curing, the wire edge features are instead limited by the
larger grain size of the annealed Au. The grains have grown from
the pre-annealed size of $50-100 \unit{nm}$ to between
$3$ and $10 \unit{\mu m}$.
Despite the enlarged grain size, the edge deviations have only increased
to about $200 \unit{nm}$.

\begin{figure}

\includegraphics{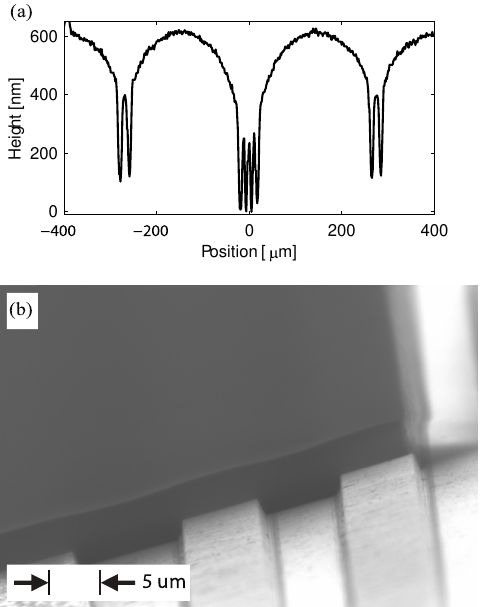}

\caption{\label{f5}
(a) Stylus profilometer (Dektak) scan of polyimide surface over the
five center wires. Polyimide was coated in three layers with full
cure between each. (b) SEM image of an etched polyimide
film demonstrating the extent of planarization over the central
trapping wires.
}
\end{figure}

The degree of planarization (DOP) of a film is defined as
DOP $=\left[ 1-\Delta h / h_{w} \right] \: \times 100\% $,
where $\Delta h$ is the deviation in the surface topography,
and $h_{w}$ is the height of the underlying features.
A single layer of PI 2562, spun at $3000 \unit{RPM}$, to a thickness
of $1.4 \unit{\mu m}$ gives
a 40\% DOP over the center wire.
Coating with two layers of PI 2562 before curing increases DOP
to 50-60\%.\ In contrast applying a layer of PI 2562 over a fully-cured
single layer gives a DOP of 70-80\%. Planarization is improved further
to 80-90\% by applying a third layer. As a demonstration of the
planarization, a region of polyimide over the center of the trapping
wires was removed using an RIE etch, as shown in Fig.~\ref{f5}b.
Wire height is $1.5 \unit{\mu m}$ and polyimide thickness is
$3.3 \unit{\mu m}$.
Figure \ref{f5}a shows a
stylus profilometer scan of the polyimide surface above the five wires.
Around the center wire, the peak-to-peak height variation of the
polyimide surface is $240 \unit{nm}$., giving a DOP of $85$\%.

The average grain size of a deposited film increases with the film
thickness and the substrate temperature during deposition
\cite{Liu:1997}. Cr/Au films were grown on polyimide surfaces at various
thicknesses to observe the effect on grain size. The temperature of the
substrates was controlled by water-cooling the substrate mount. The
average grain diameter increased from $40\unit{nm}$ for a $100\unit{nm}$
film to $60\unit{nm}$ for a $1.5 \unit{\mu m}$ film.
Magnetic field noise
caused by thermally induced stray currents increases with film
thickness -- favoring thin shield layers
\cite{zhang:2005}.

\begin{figure}
\includegraphics{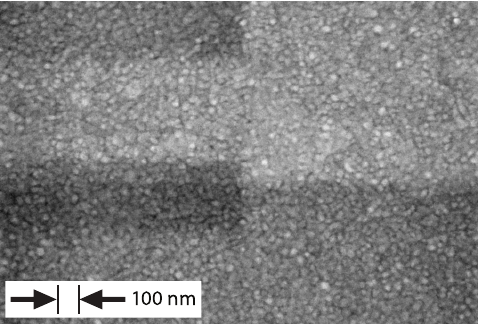}
\caption{\label{f6}
SEM image of the surface of the shield layer.
}
\end{figure}

Figure \ref{f6} shows an SEM image of the surface of the shield layer.
Average grain size is determined using a counting method, with three
smaller areas sampled from the image. The average diameter of the
grains is calculated by dividing the image area by the number of
grains counted, with the approximation that they are circular.
The average grain diameter is $40 \unit{nm}$.

\section{Discussion\label{conclusion}}

We have fabricated an atom chip for Rydberg atom-metal surface
interaction studies. The chip uses five parallel current-carrying
wires ($1.5 \unit{\mu m}$ thick Au)
over a $20 \unit{nm}$
Ti adhesion layer,
and a $50 \unit{nm}$ Pd buffer layer. The Pd is used to eliminate the
diffusion of Ti into the Au during subsequent thermal processing.
Wires are grown by thermal deposition and patterned using lift-off
photolithography.  The chip is planarized using three layers of
polyimide PI 2562 with a full cure after each layer application.
Polyimide is etched using RIE with a sputtered Al etch mask. To
avoid etching pinholes through the polyimide, the polyimide is
patterned in two stages: once after applying the first two layers
and again after the third. A total thickness of {$3.3 \unit{\mu m}$}
gives a degree of planarization of {$85$ \%}.
The chip uses a Au electrostatic shield layer above the polyimide
to isolate the Rydberg atoms from electric fields caused by potential
differences between the trapping wires. This layer is also a reflective
surface for the laser cooling beams and the surface with which the
Rydberg atoms interact. The shield is fabricated by thermal deposition
and lift-off photolithography. The Au grains of the shield are
$40 \unit{nm}$ in diameter, as measured using SEM images.

Despite their fundamental importance in a variety of experiments,
surprisingly little has been done to systematically characterize the
patch electric fields which exist over polycrystalline surfaces.
As determined in Appendix \ref{appendix:patch},
patch fields should be detectable as far as $100 \unit{\mu m}$
from our shield layer.
A careful study of these fields should be of use in the design
of experiments involving Rydberg atoms near surfaces
\cite{rensen:063601,hyafil:2004,lesanovsky:2005}.

We thank Dr.~R.~Mansour for the use of the CIRFE facility,
and R.~Al-Dahleh, B.~Jolley and
Dr.~Czang-Ho Lee for their assistance.
This work was supported by NSERC.

\begin{appendix}

\section{Experimental System Incorporating Atom Chip}
\label{appendix:experimentalsystem}

The atom chip is part of an experimental apparatus to study
Rydberg atom surface interactions.  In this appendix the entire system is described.   Some of these steps have been demonstrated by our group using a
preliminary (non-shielded) atom chip \cite{cherry:2009}.

A vapor cell mirror MOT
will cool and capture room temperature atoms of $^{87}$Rb.
After optical
molasses and optical pumping, cold atoms will be trapped using a
macroscopic Ioffe-Pritchard trap.
The cloud of atoms will then be compressed and evaporatively
cooled in this trap before transfer to the atom chip.

As far as the MOT and magnetic trapping is concerned, our approach
is conventional, following existing designs in the
literature \cite{kasper:2004,wildermuth:2004}.
A U-shaped copper wire
is embedded in the Macor beneath the chip and is used in
conjunction with a pair of Helmholtz coils to form a magnetic
quadrupole for a mirror MOT \cite{reichel:1999}.
After a sufficient number of atoms have accumulated in the MOT,
the cooling lasers are switched off, and an optical pumping beam
is introduced to transfer most of the population to the
$F=2$, $m_{F}=2$ weak-field seeking state.
They are then captured
by an intermediate magnetic trap formed by a mm-sized Z-shaped wire
(also embedded beneath the chip) and external coils. The lasers
are switched off, leaving a purely magnetic trap. Then the atoms
are evaporatively cooled using rf radiation while compressing the
trap. The cold atoms that remain are transferred to the chip by
ramping down the current in the Z-wire and increasing the currents
in the atom chip wires. Once the atoms have been loaded into the
atom chip trap, they can be positioned within a few $\mu$m of the
chip surface and optically excited to Rydberg states.

Excitation into Rydberg states can be detected by applying
an electric field pulse, which will ionize the atoms, and draw
the resulting ions or electrons
away from the surface towards a charged particle detector.
By varying the Rydberg excitation frequency, spectra may be obtained.
As discussed in Appendix \ref{appendix:patch}, the broadening
of these spectra will be used to measure patch field distributions.

\section{Control of atom surface distance}
\label{appendix:control}

The chip is suitable for confining cold clouds of atoms at distances
ranging from
$2 \unit{\mu m}$ to $200 \unit{\mu m}$ from the surface.
The inner pair of U-wires and the center wire are used to trap
the atoms at distances of less than $50 \unit{\mu m}$ from the surface.
The outer pair of U-wires and the center wire are used for larger
atom-surface distances.

\begin{figure}[t]
\includegraphics{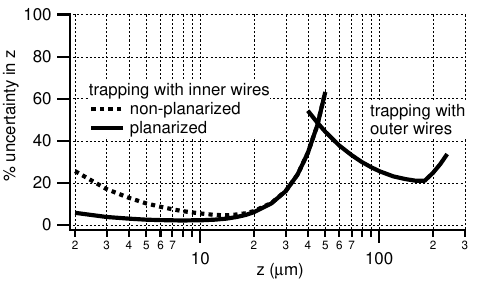}
\caption{\label{fg:control}
Uncertainty in atom-surface distance $z$
as a function of atom-surface distance
for a 10 $\mu$K cloud (see text for details).
}
\end{figure}

Two effects contribute to uncertainty in the atom-surface distance.
The cloud of atoms will have a finite size, so that not all atoms are
the same distance from the surface. Any roughness in the surface
(such as less-than-perfect planarization over the trapping wires)
will also cause variation in the atom-surface distance.
The trapping parameters at various distances were calculated,
assuming $1 \unit{\mu m}$ thick wires,
maximum current density of
$3.5 \times 10^6 \unit{A/cm^2}$,
and an applied axial ``Ioffe-Pritchard''
field of $2 \unit{G}$, large enough
to ensure that a $10 \unit{\mu K}$ cloud is
completely within the harmonic regime of the trap. The uncertainty
in the atom-surface distance due to the finite size of a
$10 \unit{\mu K}$
atom cloud alone was calculated, taking the cloud size to be defined
by the classical turning point of an atom with average kinetic energy.
The uncertainty due to surface corrugations of $1 \unit{\mu m}$
in height was
added in quadrature. Plots of the uncertainty in atom-surface distance
as a function of distance are shown in Fig.~\ref{fg:control}.
For all distances
considered, treating the $10 \unit{\mu K}$
cloud as a classical thermal cloud
is appropriate. Surface corrugations are the dominant source of uncertainty
at short distances, and the finite cloud size dominates at long distances,
where the confinement of the trap is much weaker.

At large ($> 50 \unit{\mu m}$)
atom-surface distances, a $10 \unit{\mu K}$ cloud
is undesirably large due to the weak confinement. However, the cloud
can easily be made smaller with additional rf evaporative cooling to
lower the temperature. The size of a thermal cloud in a harmonic
 potential scales as $\sqrt{T}$, so lowering the temperature of the
cloud to $1 \unit{\mu K}$
would reduce the cloud size by slightly more than a
factor of three.

As the cloud is moved toward the chip, the confinement strength
increases, so the cloud is adiabatically compressed and heated.
However, this heating can be compensated with additional rf evaporative
cooling as long as enough atoms remain for good signal-to-noise in the
Rydberg atom experiment.

\section{Patch electric fields and their detection using Rydberg
atoms}
\label{appendix:patch}

Adsorbates cause inhomogeneous electric fields above
surfaces \cite{obrecht:2007,PhysRevA.81.063411}.
However,  even in the absence of contaminants, a polycrystalline metal surface
will exhibit significant inhomogeneous electric fields at
distances from the surface extending far
beyond the atomic scale \cite{herring:1949}.
These fields are due to the differences in work function between different
crystal faces.  To estimate the size of these fields, we have extended
a simple model introduced by Rzchowski and Henderson
\cite{rzchowski:1988}.
The root-mean-square (RMS) value of the electric field is found to be \cite{carter:2011}:
\begin{equation}
E_{rms} = <\! E^2 \! >^{1/2} \: \approx
\: 0.3 \left(\frac{\Phi_{rms}}{w}\right)
\left( \frac{w}{z} \right) ^2
\end{equation}
for $z >\!> w$, where $w \times w$ is the average patch area, and $\Phi_{rms}$ is the patch to patch variation in work function (away from the mean).
For gold, $\Phi_{rms} \approx 0.1 \unit{V}$ \cite{lide:2001}.
With $w = 40 \: {\rm nm}$ for our fabricated gold shield layer, we
expect $E_{rms}=1.3 \unit{mV/cm}$
at $100 \unit{\mu m}$ from the surface.

For concreteness, we will consider the sensing of these patch
fields using the $5d_{5/2} \rightarrow 50f_{7/2}$ transition in Rb.
There are several reasons for this choice:
The complete excitation sequence
$5s \rightarrow 5p \rightarrow 5d \rightarrow 50f$
has been demonstrated previously
\cite{filipovicz:1985,nussenzveig:1993}.
The three wavelengths involved ($780$, $776$ and $1259 \unit{nm}$) are
all available from relatively inexpensive external cavity diode
laser systems, and are much longer than those required for
the $5s \rightarrow 5p \rightarrow ns$, $5s \rightarrow 5p \rightarrow nd$,
or $5s \rightarrow 5p$
excitation schemes
-- reducing photoelectron emission from the surface.
(The work function of gold can be significantly
lowered by adsorption of alkalis; see, for example,
Ref.~\cite{gray:1988}.)  In addition, compared to the $ns$, $np$
and $nd$ series, the $nf$ Rydberg states have significantly
higher polarizabilities.

\begin{figure}
\includegraphics{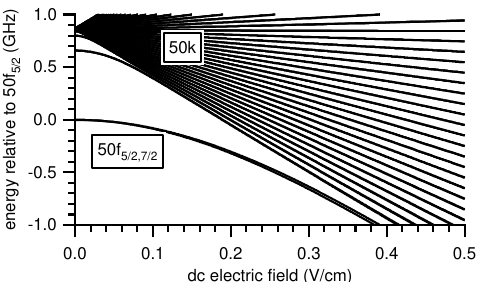}
\caption{\label{fg:stark50f}
Stark map of $m_j=1/2$ states near $50f_{5/2,7/2}$
calculated using the procedures of
Zimmerman {\it et al.} \cite{zimmerman:1979},
with zero-field spectroscopic data from Ref.~\cite{han:2006}.
}
\end{figure}

A Stark map showing the shift of the $50f$ states with electric
field is shown in Fig.~\ref{fg:stark50f}.
Due to the quadratic nature of the low
field shift it is desirable to apply a ``bias'' field to increase
spectral sensitivity to electric field.
This, however,
should not be so large that the oscillator strength
of the $nd-nf$ transition is ``diluted'' by the manifold
(the $nk$ states in Fig.~\ref{fg:stark50f}, adiabatically connected to
states of $\ell > 3$).  Dilution will reduce the signal level,
degrading the signal-to-noise (S/N).
In a field of $0.2 \unit{V/cm}$,
the state adiabatically connected with $50f_{7/2}$,
shows an electric field sensitivity of $-2.95 \unit{GHz/(V/cm)}$.
Its total $f$ character is $60 \%$, and thus, should still
be relatively straightforward to excite.

There will be contributions to the width of
the $5d_{5/2}-50f_{7/2}$ line from the finite lifetime
of the $5d$ state ($600 \unit{kHz}$ \cite{lindgard:1977}),
and the excitation laser linewidth.
With a relatively straightfoward frequency stabilization
scheme, we could expect a total linewidth of roughly
$1 \unit{MHz}$.

To avoid Zeeman broadening due to the inhomogeneous magnetic
microtrap fields, it will be necessary to turn these off shortly
prior to Rydberg state excitation.

If we assume a S/N ratio sufficient to observe an
additional $1 \unit{MHz}$ of broadening, we will be sensitive to fields
larger than
$1 \unit{MHz} / [ 2.95 \unit{GHz/(V/cm) ]}
\approx 0.3 \unit{mV/cm}$.
Thus, at $100 \unit{\mu m}$ from the shield, the patch fields
($E_{rms} = 1.3 \unit{mV/cm} $) should be observable.
As $E_{rms}$ scales like $1/z^2$, the patch fields at smaller
distances
should be observable;
although it may be desirable to use a {\em less} sensitive,
lower-$n$ Rydberg state closer to the surface to avoid
spectral confusion.

Rydberg atoms near metal surfaces show an energy shift due
to a dipole-dipole interaction with their
electrostatic images (the Lennard-Jones shift \cite{hinds:1991}).
This shift is split due to magnetic sublevel
structure, and contributes to broadening
due to the variation in atom-surface
distance discussed in Appendix \ref{appendix:control}.
Therefore, to sense the patch fields, the Lennard
Jones shift should be less than the broadening due to the patch fields.
Using the formulae of Ref.~\cite{hinds:1991}
we estimate the Lennard-Jones shift of the
$50f_{7/2}$ $m_j=1/2$ state -- the axis of quantization
is the surface normal -- at
$z= 100 \unit{\mu m}$ to be $1.4 \unit{kHz}$.
(This is probably an overestimate, as it assumes a perfectly
conducting surface.)
This shift scales like $n^4/z^3$ \cite{sandoghdar:1992}
-- at $10 \unit {\mu m}$
it is $1.4 \unit{MHz}$.
However, this is still much smaller than the expected Stark broadening
due to patch fields.  A combination of zero average dc field,
together with low $z$ and $n$ is best
for enhancing the Lennard-Jones shift relative to
Stark broadening \cite{sandoghdar:1992}.

\end{appendix}

\bibliography{references}

\begin{thebibliography}{40}%
\makeatletter
\providecommand \@ifxundefined [1]{%
 \@ifx{#1\undefined}
}%
\providecommand \@ifnum [1]{%
 \ifnum #1\expandafter \@firstoftwo
 \else \expandafter \@secondoftwo
 \fi
}%
\providecommand \@ifx [1]{%
 \ifx #1\expandafter \@firstoftwo
 \else \expandafter \@secondoftwo
 \fi
}%
\providecommand \natexlab [1]{#1}%
\providecommand \enquote  [1]{``#1''}%
\providecommand \bibnamefont  [1]{#1}%
\providecommand \bibfnamefont [1]{#1}%
\providecommand \citenamefont [1]{#1}%
\providecommand \href@noop [0]{\@secondoftwo}%
\providecommand \href [0]{\begingroup \@sanitize@url \@href}%
\providecommand \@href[1]{\@@startlink{#1}\@@href}%
\providecommand \@@href[1]{\endgroup#1\@@endlink}%
\providecommand \@sanitize@url [0]{\catcode `\\12\catcode `\$12\catcode
  `\&12\catcode `\#12\catcode `\^12\catcode `\_12\catcode `\%12\relax}%
\providecommand \@@startlink[1]{}%
\providecommand \@@endlink[0]{}%
\providecommand \url  [0]{\begingroup\@sanitize@url \@url }%
\providecommand \@url [1]{\endgroup\@href {#1}{\urlprefix }}%
\providecommand \urlprefix  [0]{URL }%
\providecommand \Eprint [0]{\href }%
\providecommand \doibase [0]{http://dx.doi.org/}%
\providecommand \selectlanguage [0]{\@gobble}%
\providecommand \bibinfo  [0]{\@secondoftwo}%
\providecommand \bibfield  [0]{\@secondoftwo}%
\providecommand \translation [1]{[#1]}%
\providecommand \BibitemOpen [0]{}%
\providecommand \bibitemStop [0]{}%
\providecommand \bibitemNoStop [0]{.\EOS\space}%
\providecommand \EOS [0]{\spacefactor3000\relax}%
\providecommand \BibitemShut  [1]{\csname bibitem#1\endcsname}%
\let\auto@bib@innerbib\@empty
\bibitem [{\citenamefont {Gallagher}(1994)}]{gallagher:1994}%
  \BibitemOpen
  \bibfield  {author} {\bibinfo {author} {\bibfnamefont {T.~F.}\ \bibnamefont
  {Gallagher}},\ }\href@noop {} {\emph {\bibinfo {title} {Rydberg Atoms}}}\
  (\bibinfo  {publisher} {Cambridge University Press},\ \bibinfo {address}
  {Cambridge},\ \bibinfo {year} {1994})\BibitemShut {NoStop}%
\bibitem [{\citenamefont {Chaplik}(1968)}]{chaplik:1968}%
  \BibitemOpen
  \bibfield  {author} {\bibinfo {author} {\bibfnamefont {A.~V.}\ \bibnamefont
  {Chaplik}},\ }\href@noop {} {\bibfield  {journal} {\bibinfo  {journal}
  {Soviet Physics JETP}\ }\textbf {\bibinfo {volume} {27}},\ \bibinfo {pages}
  {178} (\bibinfo {year} {1968})}\BibitemShut {NoStop}%
\bibitem [{\citenamefont {Sandoghdar}\ \emph {et~al.}(1992)\citenamefont
  {Sandoghdar}, \citenamefont {Sukenik}, \citenamefont {Hinds},\ and\
  \citenamefont {Haroche}}]{sandoghdar:1992}%
  \BibitemOpen
  \bibfield  {author} {\bibinfo {author} {\bibfnamefont {V.}~\bibnamefont
  {Sandoghdar}}, \bibinfo {author} {\bibfnamefont {C.~I.}\ \bibnamefont
  {Sukenik}}, \bibinfo {author} {\bibfnamefont {E.~A.}\ \bibnamefont {Hinds}},
  \ and\ \bibinfo {author} {\bibfnamefont {S.}~\bibnamefont {Haroche}},\
  }\href@noop {} {\bibfield  {journal} {\bibinfo  {journal} {Phys. Rev. Lett.}\
  }\textbf {\bibinfo {volume} {68}},\ \bibinfo {pages} {3432} (\bibinfo {year}
  {1992})}\BibitemShut {NoStop}%
\bibitem [{\citenamefont {Hill}\ \emph {et~al.}(2000)\citenamefont {Hill},
  \citenamefont {Haich}, \citenamefont {Zhou}, \citenamefont {Nordlander},\
  and\ \citenamefont {Dunning}}]{hill:2000}%
  \BibitemOpen
  \bibfield  {author} {\bibinfo {author} {\bibfnamefont {S.~B.}\ \bibnamefont
  {Hill}}, \bibinfo {author} {\bibfnamefont {C.~B.}\ \bibnamefont {Haich}},
  \bibinfo {author} {\bibfnamefont {Z.}~\bibnamefont {Zhou}}, \bibinfo {author}
  {\bibfnamefont {P.}~\bibnamefont {Nordlander}}, \ and\ \bibinfo {author}
  {\bibfnamefont {F.~B.}\ \bibnamefont {Dunning}},\ }\href@noop {} {\bibfield
  {journal} {\bibinfo  {journal} {Phys. Rev. Lett.}\ }\textbf {\bibinfo
  {volume} {85}},\ \bibinfo {pages} {5444} (\bibinfo {year}
  {2000})}\BibitemShut {NoStop}%
\bibitem [{\citenamefont {Folman}\ \emph {et~al.}(2002)\citenamefont {Folman},
  \citenamefont {Kr\"{u}ger}, \citenamefont {Schmiedmayer}, \citenamefont
  {Denschlag},\ and\ \citenamefont {Henkel}}]{folman:2002}%
  \BibitemOpen
  \bibfield  {author} {\bibinfo {author} {\bibfnamefont {R.}~\bibnamefont
  {Folman}}, \bibinfo {author} {\bibfnamefont {P.}~\bibnamefont {Kr\"{u}ger}},
  \bibinfo {author} {\bibfnamefont {J.}~\bibnamefont {Schmiedmayer}}, \bibinfo
  {author} {\bibfnamefont {J.}~\bibnamefont {Denschlag}}, \ and\ \bibinfo
  {author} {\bibfnamefont {C.}~\bibnamefont {Henkel}},\ }\href@noop {}
  {\bibfield  {journal} {\bibinfo  {journal} {Adv. At. Mol. Opt. Phys.}\
  }\textbf {\bibinfo {volume} {48}},\ \bibinfo {pages} {263} (\bibinfo {year}
  {2002})}\BibitemShut {NoStop}%
\bibitem [{\citenamefont {Fort\'{a}gh}\ and\ \citenamefont
  {Zimmerman}(2007)}]{fortagh:2007}%
  \BibitemOpen
  \bibfield  {author} {\bibinfo {author} {\bibfnamefont {J.}~\bibnamefont
  {Fort\'{a}gh}}\ and\ \bibinfo {author} {\bibfnamefont {C.}~\bibnamefont
  {Zimmerman}},\ }\href@noop {} {\bibfield  {journal} {\bibinfo  {journal}
  {Rev. Mod. Phys.}\ }\textbf {\bibinfo {volume} {79}},\ \bibinfo {pages} {235}
  (\bibinfo {year} {2007})}\BibitemShut {NoStop}%
\bibitem [{\citenamefont {Lin}\ \emph {et~al.}(2004)\citenamefont {Lin},
  \citenamefont {Teper}, \citenamefont {Chin},\ and\ \citenamefont
  {Vuleti\'{c}}}]{lin:2004}%
  \BibitemOpen
  \bibfield  {author} {\bibinfo {author} {\bibfnamefont {Y.-J.}\ \bibnamefont
  {Lin}}, \bibinfo {author} {\bibfnamefont {I.}~\bibnamefont {Teper}}, \bibinfo
  {author} {\bibfnamefont {C.}~\bibnamefont {Chin}}, \ and\ \bibinfo {author}
  {\bibfnamefont {V.}~\bibnamefont {Vuleti\'{c}}},\ }\href@noop {} {\bibfield
  {journal} {\bibinfo  {journal} {Phys. Rev. Lett.}\ }\textbf {\bibinfo
  {volume} {92}},\ \bibinfo {pages} {050404} (\bibinfo {year}
  {2004})}\BibitemShut {NoStop}%
\bibitem [{\citenamefont {Tauschinsky}\ \emph {et~al.}(2010)\citenamefont
  {Tauschinsky}, \citenamefont {Thijssen}, \citenamefont {Whitlock},
  \citenamefont {van Linden van~den Heuvell},\ and\ \citenamefont
  {Spreeuw}}]{PhysRevA.81.063411}%
  \BibitemOpen
  \bibfield  {author} {\bibinfo {author} {\bibfnamefont {A.}~\bibnamefont
  {Tauschinsky}}, \bibinfo {author} {\bibfnamefont {R.~M.~T.}\ \bibnamefont
  {Thijssen}}, \bibinfo {author} {\bibfnamefont {S.}~\bibnamefont {Whitlock}},
  \bibinfo {author} {\bibfnamefont {H.~B.}\ \bibnamefont {van Linden van~den
  Heuvell}}, \ and\ \bibinfo {author} {\bibfnamefont {R.~J.~C.}\ \bibnamefont
  {Spreeuw}},\ }\href {\doibase 10.1103/PhysRevA.81.063411} {\bibfield
  {journal} {\bibinfo  {journal} {Phys. Rev. A}\ }\textbf {\bibinfo {volume}
  {81}},\ \bibinfo {pages} {063411} (\bibinfo {year} {2010})}\BibitemShut
  {NoStop}%
\bibitem [{\citenamefont {Reichel}\ \emph {et~al.}(1999)\citenamefont
  {Reichel}, \citenamefont {H\"{a}nsel},\ and\ \citenamefont
  {H\"{a}nsch}}]{reichel:1999}%
  \BibitemOpen
  \bibfield  {author} {\bibinfo {author} {\bibfnamefont {J.}~\bibnamefont
  {Reichel}}, \bibinfo {author} {\bibfnamefont {W.}~\bibnamefont {H\"{a}nsel}},
  \ and\ \bibinfo {author} {\bibfnamefont {T.}~\bibnamefont {H\"{a}nsch}},\
  }\href@noop {} {\bibfield  {journal} {\bibinfo  {journal} {Phys. Rev. Lett.}\
  }\textbf {\bibinfo {volume} {83}},\ \bibinfo {pages} {3398} (\bibinfo {year}
  {1999})}\BibitemShut {NoStop}%
\bibitem [{\citenamefont {Reichel}\ \emph {et~al.}(2001)\citenamefont
  {Reichel}, \citenamefont {H\"{a}nsel}, \citenamefont {Hommelhoff},\ and\
  \citenamefont {H\"{a}nsch}}]{reichel:2001}%
  \BibitemOpen
  \bibfield  {author} {\bibinfo {author} {\bibfnamefont {J.}~\bibnamefont
  {Reichel}}, \bibinfo {author} {\bibfnamefont {W.}~\bibnamefont {H\"{a}nsel}},
  \bibinfo {author} {\bibnamefont {Hommelhoff}}, \ and\ \bibinfo {author}
  {\bibfnamefont {T.}~\bibnamefont {H\"{a}nsch}},\ }\href@noop {} {\bibfield
  {journal} {\bibinfo  {journal} {Appl. Phys. B}\ }\textbf {\bibinfo {volume}
  {72}},\ \bibinfo {pages} {81} (\bibinfo {year} {2001})}\BibitemShut {NoStop}%
\bibitem [{\citenamefont {Lev}(2006)}]{thesis:lev}%
  \BibitemOpen
  \bibfield  {author} {\bibinfo {author} {\bibfnamefont {B.}~\bibnamefont
  {Lev}},\ }\emph {\bibinfo {title} {Magnetic microtraps for cavity {QED},
  {B}ose-{E}instein condensates, and atom optics}},\ \href@noop {} {Ph.D.
  thesis},\ \bibinfo  {school} {California Institute of Technology} (\bibinfo
  {year} {2006})\BibitemShut {NoStop}%
\bibitem [{\citenamefont {Treutlein}\ \emph {et~al.}(2006)\citenamefont
  {Treutlein}, \citenamefont {H\"{a}nsch}, \citenamefont {Reichel},
  \citenamefont {Negretti}, \citenamefont {Cirone},\ and\ \citenamefont
  {Calarco}}]{treutlein:2006}%
  \BibitemOpen
  \bibfield  {author} {\bibinfo {author} {\bibfnamefont {P.}~\bibnamefont
  {Treutlein}}, \bibinfo {author} {\bibfnamefont {T.}~\bibnamefont
  {H\"{a}nsch}}, \bibinfo {author} {\bibfnamefont {J.}~\bibnamefont {Reichel}},
  \bibinfo {author} {\bibfnamefont {A.}~\bibnamefont {Negretti}}, \bibinfo
  {author} {\bibfnamefont {M.}~\bibnamefont {Cirone}}, \ and\ \bibinfo {author}
  {\bibfnamefont {T.}~\bibnamefont {Calarco}},\ }\href@noop {} {\bibfield
  {journal} {\bibinfo  {journal} {Phys. Rev. A}\ }\textbf {\bibinfo {volume}
  {74}},\ \bibinfo {pages} {022312} (\bibinfo {year} {2006})}\BibitemShut
  {NoStop}%
\bibitem [{\citenamefont {Nirrengarten}\ \emph {et~al.}(2006)\citenamefont
  {Nirrengarten}, \citenamefont {Qarry}, \citenamefont {Roux}, \citenamefont
  {Emmert}, \citenamefont {Nogues}, \citenamefont {Brune}, \citenamefont
  {Raimond},\ and\ \citenamefont
  {Haroche}}]{article:harochesuperconductingchip}%
  \BibitemOpen
  \bibfield  {author} {\bibinfo {author} {\bibfnamefont {T.}~\bibnamefont
  {Nirrengarten}}, \bibinfo {author} {\bibfnamefont {A.}~\bibnamefont {Qarry}},
  \bibinfo {author} {\bibfnamefont {C.}~\bibnamefont {Roux}}, \bibinfo {author}
  {\bibfnamefont {A.}~\bibnamefont {Emmert}}, \bibinfo {author} {\bibfnamefont
  {G.}~\bibnamefont {Nogues}}, \bibinfo {author} {\bibfnamefont
  {M.}~\bibnamefont {Brune}}, \bibinfo {author} {\bibfnamefont {J.-M.}\
  \bibnamefont {Raimond}}, \ and\ \bibinfo {author} {\bibfnamefont
  {S.}~\bibnamefont {Haroche}},\ }\href@noop {} {\bibfield  {journal} {\bibinfo
   {journal} {Phys. Rev. Lett.}\ }\textbf {\bibinfo {volume} {97}},\ \bibinfo
  {pages} {200405} (\bibinfo {year} {2006})}\BibitemShut {NoStop}%
\bibitem [{\citenamefont {Est\`{e}ve}\ \emph {et~al.}(2004)\citenamefont
  {Est\`{e}ve}, \citenamefont {Aussibal}, \citenamefont {Schumm}, \citenamefont
  {Figl}, \citenamefont {Mailly}, \citenamefont {Bouchoule}, \citenamefont
  {Westbrook},\ and\ \citenamefont {Aspect}}]{article:estevecorrugation2}%
  \BibitemOpen
  \bibfield  {author} {\bibinfo {author} {\bibfnamefont {J.}~\bibnamefont
  {Est\`{e}ve}}, \bibinfo {author} {\bibfnamefont {C.}~\bibnamefont
  {Aussibal}}, \bibinfo {author} {\bibfnamefont {T.}~\bibnamefont {Schumm}},
  \bibinfo {author} {\bibfnamefont {C.}~\bibnamefont {Figl}}, \bibinfo {author}
  {\bibfnamefont {D.}~\bibnamefont {Mailly}}, \bibinfo {author} {\bibfnamefont
  {I.}~\bibnamefont {Bouchoule}}, \bibinfo {author} {\bibfnamefont
  {C.}~\bibnamefont {Westbrook}}, \ and\ \bibinfo {author} {\bibfnamefont
  {A.}~\bibnamefont {Aspect}},\ }\href@noop {} {\bibfield  {journal} {\bibinfo
  {journal} {Phys. Rev. A}\ }\textbf {\bibinfo {volume} {70}},\ \bibinfo
  {pages} {043629} (\bibinfo {year} {2004})}\BibitemShut {NoStop}%
\bibitem [{\citenamefont {Groth}\ \emph {et~al.}(2004)\citenamefont {Groth},
  \citenamefont {Kr\"{u}ger}, \citenamefont {Wildermuth}, \citenamefont
  {Folman}, \citenamefont {Fernholz},\ and\ \citenamefont
  {Schmiedmayer}}]{groth:2004}%
  \BibitemOpen
  \bibfield  {author} {\bibinfo {author} {\bibfnamefont {S.}~\bibnamefont
  {Groth}}, \bibinfo {author} {\bibfnamefont {P.}~\bibnamefont {Kr\"{u}ger}},
  \bibinfo {author} {\bibfnamefont {S.}~\bibnamefont {Wildermuth}}, \bibinfo
  {author} {\bibfnamefont {R.}~\bibnamefont {Folman}}, \bibinfo {author}
  {\bibfnamefont {T.}~\bibnamefont {Fernholz}}, \ and\ \bibinfo {author}
  {\bibfnamefont {J.}~\bibnamefont {Schmiedmayer}},\ }\href@noop {} {\bibfield
  {journal} {\bibinfo  {journal} {Appl. Phys. Lett.}\ }\textbf {\bibinfo
  {volume} {85}},\ \bibinfo {pages} {2980} (\bibinfo {year}
  {2004})}\BibitemShut {NoStop}%
\bibitem [{\citenamefont {Kern}\ and\ \citenamefont
  {Puotinen}(1970)}]{kern:1970}%
  \BibitemOpen
  \bibfield  {author} {\bibinfo {author} {\bibfnamefont {W.}~\bibnamefont
  {Kern}}\ and\ \bibinfo {author} {\bibfnamefont {D.}~\bibnamefont
  {Puotinen}},\ }\href@noop {} {\bibfield  {journal} {\bibinfo  {journal} {RCA
  Rev.}\ ,\ \bibinfo {pages} {187}} (\bibinfo {year} {1970})}\BibitemShut
  {NoStop}%
\bibitem [{dat(2007)}]{datasheet:AZ_2000}%
  \BibitemOpen
  \href@noop {} {\emph {\bibinfo {title} {A{Z} \textit{n}{LOF} 2000
  {P}hotoresist {D}ata {S}heet}}},\ \bibinfo {organization} {{C}lariant-{AZ}
  {El}ectronic {M}aterials} (\bibinfo {year} {2007})\BibitemShut {NoStop}%
\bibitem [{\citenamefont {Madakson}(1991)}]{article:madakson_diffusion}%
  \BibitemOpen
  \bibfield  {author} {\bibinfo {author} {\bibfnamefont {P.}~\bibnamefont
  {Madakson}},\ }\href@noop {} {\bibfield  {journal} {\bibinfo  {journal} {J.
  Appl. Phys.}\ }\textbf {\bibinfo {volume} {79}},\ \bibinfo {pages} {1380}
  (\bibinfo {year} {1991})}\BibitemShut {NoStop}%
\bibitem [{\citenamefont {Munitz}\ and\ \citenamefont
  {Komem}(1976)}]{Munitz:diffusion}%
  \BibitemOpen
  \bibfield  {author} {\bibinfo {author} {\bibfnamefont {A.}~\bibnamefont
  {Munitz}}\ and\ \bibinfo {author} {\bibfnamefont {Y.}~\bibnamefont {Komem}},\
  }\href@noop {} {\bibfield  {journal} {\bibinfo  {journal} {Thin Solid Films}\
  }\textbf {\bibinfo {volume} {37}},\ \bibinfo {pages} {171} (\bibinfo {year}
  {1976})}\BibitemShut {NoStop}%
\bibitem [{\citenamefont {Pan}\ \emph {et~al.}(2004)\citenamefont {Pan},
  \citenamefont {Pafchek}, \citenamefont {Judd},\ and\ \citenamefont
  {Baxter}}]{conference:Pan_diffusion}%
  \BibitemOpen
  \bibfield  {author} {\bibinfo {author} {\bibfnamefont {J.}~\bibnamefont
  {Pan}}, \bibinfo {author} {\bibfnamefont {M.}~\bibnamefont {Pafchek}},
  \bibinfo {author} {\bibfnamefont {F.}~\bibnamefont {Judd}}, \ and\ \bibinfo
  {author} {\bibfnamefont {J.}~\bibnamefont {Baxter}}\ }(\bibinfo
  {organization} {IEEE/SEMI International Elecronics Manufacturing Technology
  Symposium},\ \bibinfo {year} {2004})\BibitemShut {NoStop}%
\bibitem [{\citenamefont {Liu}\ \emph {et~al.}(1997)\citenamefont {Liu},
  \citenamefont {Brown},\ and\ \citenamefont {McKinley}}]{Liu:1997}%
  \BibitemOpen
  \bibfield  {author} {\bibinfo {author} {\bibfnamefont {Z.}~\bibnamefont
  {Liu}}, \bibinfo {author} {\bibfnamefont {N.}~\bibnamefont {Brown}}, \ and\
  \bibinfo {author} {\bibfnamefont {A.}~\bibnamefont {McKinley}},\ }\href@noop
  {} {\bibfield  {journal} {\bibinfo  {journal} {J. Phys. Cond. Matt.}\
  }\textbf {\bibinfo {volume} {9}},\ \bibinfo {pages} {59} (\bibinfo {year}
  {1997})}\BibitemShut {NoStop}%
\bibitem [{\citenamefont {Zhang}\ \emph {et~al.}(2005)\citenamefont {Zhang},
  \citenamefont {Henkel}, \citenamefont {Haller}, \citenamefont {Wildermuth},
  \citenamefont {Hofferberth}, \citenamefont {Kr\"{u}ger},\ and\ \citenamefont
  {Schmiedmayer}}]{zhang:2005}%
  \BibitemOpen
  \bibfield  {author} {\bibinfo {author} {\bibfnamefont {B.}~\bibnamefont
  {Zhang}}, \bibinfo {author} {\bibfnamefont {C.}~\bibnamefont {Henkel}},
  \bibinfo {author} {\bibfnamefont {E.}~\bibnamefont {Haller}}, \bibinfo
  {author} {\bibfnamefont {S.}~\bibnamefont {Wildermuth}}, \bibinfo {author}
  {\bibfnamefont {S.}~\bibnamefont {Hofferberth}}, \bibinfo {author}
  {\bibfnamefont {P.}~\bibnamefont {Kr\"{u}ger}}, \ and\ \bibinfo {author}
  {\bibfnamefont {J.}~\bibnamefont {Schmiedmayer}},\ }\href@noop {} {\bibfield
  {journal} {\bibinfo  {journal} {Eur. Phys. J. D}\ }\textbf {\bibinfo {volume}
  {35}},\ \bibinfo {pages} {97} (\bibinfo {year} {2005})}\BibitemShut {NoStop}%
\bibitem [{\citenamefont {{S\o rensen}}\ \emph {et~al.}(2004)\citenamefont
  {{S\o rensen}}, \citenamefont {van~der Wal}, \citenamefont {Childress},\ and\
  \citenamefont {Lukin}}]{rensen:063601}%
  \BibitemOpen
  \bibfield  {author} {\bibinfo {author} {\bibfnamefont {A.~S.}\ \bibnamefont
  {{S\o rensen}}}, \bibinfo {author} {\bibfnamefont {C.~H.}\ \bibnamefont
  {van~der Wal}}, \bibinfo {author} {\bibfnamefont {L.~I.}\ \bibnamefont
  {Childress}}, \ and\ \bibinfo {author} {\bibfnamefont {M.~D.}\ \bibnamefont
  {Lukin}},\ }\href@noop {} {\bibfield  {journal} {\bibinfo  {journal} {Phys.
  Rev. Lett.}\ }\textbf {\bibinfo {volume} {92}},\ \bibinfo {eid} {063601}
  (\bibinfo {year} {2004})}\BibitemShut {NoStop}%
\bibitem [{\citenamefont {Hyafil}\ \emph {et~al.}(2004)\citenamefont {Hyafil},
  \citenamefont {Mozley}, \citenamefont {Perrin}, \citenamefont {Tailleur},
  \citenamefont {Nogues}, \citenamefont {Brune}, \citenamefont {Raimond},\ and\
  \citenamefont {Haroche}}]{hyafil:2004}%
  \BibitemOpen
  \bibfield  {author} {\bibinfo {author} {\bibfnamefont {P.}~\bibnamefont
  {Hyafil}}, \bibinfo {author} {\bibfnamefont {J.}~\bibnamefont {Mozley}},
  \bibinfo {author} {\bibfnamefont {A.}~\bibnamefont {Perrin}}, \bibinfo
  {author} {\bibfnamefont {J.}~\bibnamefont {Tailleur}}, \bibinfo {author}
  {\bibfnamefont {G.}~\bibnamefont {Nogues}}, \bibinfo {author} {\bibfnamefont
  {M.}~\bibnamefont {Brune}}, \bibinfo {author} {\bibfnamefont {J.-M.}\
  \bibnamefont {Raimond}}, \ and\ \bibinfo {author} {\bibfnamefont
  {S.}~\bibnamefont {Haroche}},\ }\href@noop {} {\bibfield  {journal} {\bibinfo
   {journal} {Phys. Rev. Lett.}\ }\textbf {\bibinfo {volume} {93}},\ \bibinfo
  {pages} {103001} (\bibinfo {year} {2004})}\BibitemShut {NoStop}%
\bibitem [{\citenamefont {Lesanovsky}\ and\ \citenamefont
  {Schmelcher}(2005)}]{lesanovsky:2005}%
  \BibitemOpen
  \bibfield  {author} {\bibinfo {author} {\bibfnamefont {I.}~\bibnamefont
  {Lesanovsky}}\ and\ \bibinfo {author} {\bibfnamefont {P.}~\bibnamefont
  {Schmelcher}},\ }\href@noop {} {\bibfield  {journal} {\bibinfo  {journal}
  {Phys. Rev. Lett.}\ }\textbf {\bibinfo {volume} {95}},\ \bibinfo {pages}
  {53001} (\bibinfo {year} {2005})}\BibitemShut {NoStop}%
\bibitem [{\citenamefont {Cherry}\ \emph {et~al.}(2009)\citenamefont {Cherry},
  \citenamefont {Carter},\ and\ \citenamefont {Martin}}]{cherry:2009}%
  \BibitemOpen
  \bibfield  {author} {\bibinfo {author} {\bibfnamefont {O.}~\bibnamefont
  {Cherry}}, \bibinfo {author} {\bibfnamefont {J.~D.}\ \bibnamefont {Carter}},
  \ and\ \bibinfo {author} {\bibfnamefont {J.~D.~D.}\ \bibnamefont {Martin}},\
  }\href@noop {} {\bibfield  {journal} {\bibinfo  {journal} {Can. J. Phys.}\
  }\textbf {\bibinfo {volume} {87}},\ \bibinfo {pages} {633} (\bibinfo {year}
  {2009})}\BibitemShut {NoStop}%
\bibitem [{\citenamefont {Kasper}\ \emph {et~al.}(2004)\citenamefont {Kasper},
  \citenamefont {Schneider}, \citenamefont {vom Hagen}, \citenamefont
  {Bartenstein}, \citenamefont {Engeser}, \citenamefont {Schumm}, \citenamefont
  {Bar-Joseph}, \citenamefont {Folman}, \citenamefont {Feenstra},\ and\
  \citenamefont {Schmiedmayer}}]{kasper:2004}%
  \BibitemOpen
  \bibfield  {author} {\bibinfo {author} {\bibfnamefont {A.}~\bibnamefont
  {Kasper}}, \bibinfo {author} {\bibfnamefont {S.}~\bibnamefont {Schneider}},
  \bibinfo {author} {\bibfnamefont {C.}~\bibnamefont {vom Hagen}}, \bibinfo
  {author} {\bibfnamefont {M.}~\bibnamefont {Bartenstein}}, \bibinfo {author}
  {\bibfnamefont {B.}~\bibnamefont {Engeser}}, \bibinfo {author} {\bibfnamefont
  {T.}~\bibnamefont {Schumm}}, \bibinfo {author} {\bibfnamefont
  {I.}~\bibnamefont {Bar-Joseph}}, \bibinfo {author} {\bibfnamefont
  {R.}~\bibnamefont {Folman}}, \bibinfo {author} {\bibfnamefont
  {L.}~\bibnamefont {Feenstra}}, \ and\ \bibinfo {author} {\bibfnamefont
  {J.}~\bibnamefont {Schmiedmayer}},\ }\href@noop {} {\bibfield  {journal}
  {\bibinfo  {journal} {J. Opt. B: Quant. Semiclass. Opt}\ }\textbf {\bibinfo
  {volume} {85}},\ \bibinfo {pages} {2980} (\bibinfo {year}
  {2004})}\BibitemShut {NoStop}%
\bibitem [{\citenamefont {Wildermuth}\ \emph {et~al.}(2004)\citenamefont
  {Wildermuth}, \citenamefont {Kr\"{u}ger}, \citenamefont {Becker},
  \citenamefont {Brajdic}, \citenamefont {Haupt}, \citenamefont {Kasper},
  \citenamefont {Folman},\ and\ \citenamefont
  {Schmiedmayer}}]{wildermuth:2004}%
  \BibitemOpen
  \bibfield  {author} {\bibinfo {author} {\bibfnamefont {S.}~\bibnamefont
  {Wildermuth}}, \bibinfo {author} {\bibfnamefont {P.}~\bibnamefont
  {Kr\"{u}ger}}, \bibinfo {author} {\bibfnamefont {C.}~\bibnamefont {Becker}},
  \bibinfo {author} {\bibfnamefont {M.}~\bibnamefont {Brajdic}}, \bibinfo
  {author} {\bibfnamefont {S.}~\bibnamefont {Haupt}}, \bibinfo {author}
  {\bibfnamefont {A.}~\bibnamefont {Kasper}}, \bibinfo {author} {\bibfnamefont
  {R.}~\bibnamefont {Folman}}, \ and\ \bibinfo {author} {\bibfnamefont
  {J.}~\bibnamefont {Schmiedmayer}},\ }\href@noop {} {\bibfield  {journal}
  {\bibinfo  {journal} {Phys. Rev. A}\ }\textbf {\bibinfo {volume} {69}},\
  \bibinfo {pages} {030901} (\bibinfo {year} {2004})}\BibitemShut {NoStop}%
\bibitem [{\citenamefont {Obrecht}\ \emph {et~al.}(2007)\citenamefont
  {Obrecht}, \citenamefont {Wild},\ and\ \citenamefont
  {Cornell}}]{obrecht:2007}%
  \BibitemOpen
  \bibfield  {author} {\bibinfo {author} {\bibfnamefont {J.~M.}\ \bibnamefont
  {Obrecht}}, \bibinfo {author} {\bibfnamefont {R.~J.}\ \bibnamefont {Wild}}, \
  and\ \bibinfo {author} {\bibfnamefont {E.~A.}\ \bibnamefont {Cornell}},\
  }\href@noop {} {\bibfield  {journal} {\bibinfo  {journal} {Phys. Rev. A}\
  }\textbf {\bibinfo {volume} {75}},\ \bibinfo {pages} {062903} (\bibinfo
  {year} {2007})}\BibitemShut {NoStop}%
\bibitem [{\citenamefont {Herring}\ and\ \citenamefont
  {Nichols}(1949)}]{herring:1949}%
  \BibitemOpen
  \bibfield  {author} {\bibinfo {author} {\bibfnamefont {C.}~\bibnamefont
  {Herring}}\ and\ \bibinfo {author} {\bibfnamefont {M.~H.}\ \bibnamefont
  {Nichols}},\ }\href@noop {} {\bibfield  {journal} {\bibinfo  {journal} {Rev.
  Mod. Phys.}\ }\textbf {\bibinfo {volume} {21}},\ \bibinfo {pages} {185}
  (\bibinfo {year} {1949})}\BibitemShut {NoStop}%
\bibitem [{\citenamefont {Rzchowski}\ and\ \citenamefont
  {Henderson}(1988)}]{rzchowski:1988}%
  \BibitemOpen
  \bibfield  {author} {\bibinfo {author} {\bibfnamefont {M.~S.}\ \bibnamefont
  {Rzchowski}}\ and\ \bibinfo {author} {\bibfnamefont {J.~R.}\ \bibnamefont
  {Henderson}},\ }\href@noop {} {\bibfield  {journal} {\bibinfo  {journal}
  {Phys. Rev. A}\ }\textbf {\bibinfo {volume} {38}},\ \bibinfo {pages} {4622}
  (\bibinfo {year} {1988})}\BibitemShut {NoStop}%
\bibitem [{\citenamefont {Carter}\ and\ \citenamefont
  {Martin}(2011)}]{carter:2011}%
  \BibitemOpen
  \bibfield  {author} {\bibinfo {author} {\bibfnamefont {J.~D.}\ \bibnamefont
  {Carter}}\ and\ \bibinfo {author} {\bibfnamefont {J.~D.~D.}\ \bibnamefont
  {Martin}},\ }\href@noop {} {\bibfield  {journal} {\bibinfo  {journal} {Phys.
  Rev. A}\ }\textbf {\bibinfo {volume} {83}},\ \bibinfo {pages} {032902}
  (\bibinfo {year} {2011})}\BibitemShut {NoStop}%
\bibitem [{\citenamefont {Lide}(2001)}]{lide:2001}%
  \BibitemOpen
  \bibinfo {editor} {\bibfnamefont {C.~R.}\ \bibnamefont {Lide}},\ ed.,\
  \href@noop {} {\emph {\bibinfo {title} {CRC Handbook of Chemistry and
  Physics}}},\ \bibinfo {edition} {82nd}\ ed.\ (\bibinfo  {publisher} {CRC
  Press},\ \bibinfo {address} {Boca Raton, FL},\ \bibinfo {year}
  {2001})\BibitemShut {NoStop}%
\bibitem [{\citenamefont {Filipovicz}\ \emph {et~al.}(1985)\citenamefont
  {Filipovicz}, \citenamefont {Meystre}, \citenamefont {Rempe},\ and\
  \citenamefont {Walther}}]{filipovicz:1985}%
  \BibitemOpen
  \bibfield  {author} {\bibinfo {author} {\bibfnamefont {P.}~\bibnamefont
  {Filipovicz}}, \bibinfo {author} {\bibfnamefont {P.}~\bibnamefont {Meystre}},
  \bibinfo {author} {\bibfnamefont {G.}~\bibnamefont {Rempe}}, \ and\ \bibinfo
  {author} {\bibfnamefont {H.}~\bibnamefont {Walther}},\ }\href@noop {}
  {\bibfield  {journal} {\bibinfo  {journal} {Opt. Acta}\ }\textbf {\bibinfo
  {volume} {32}},\ \bibinfo {pages} {1105} (\bibinfo {year}
  {1985})}\BibitemShut {NoStop}%
\bibitem [{\citenamefont {Nussenzveig}\ \emph {et~al.}(1993)\citenamefont
  {Nussenzveig}, \citenamefont {Bernardot}, \citenamefont {Brune},
  \citenamefont {Hare}, \citenamefont {Raimond}, \citenamefont {Haroche},\ and\
  \citenamefont {Gawlik}}]{nussenzveig:1993}%
  \BibitemOpen
  \bibfield  {author} {\bibinfo {author} {\bibfnamefont {P.}~\bibnamefont
  {Nussenzveig}}, \bibinfo {author} {\bibfnamefont {F.}~\bibnamefont
  {Bernardot}}, \bibinfo {author} {\bibfnamefont {M.}~\bibnamefont {Brune}},
  \bibinfo {author} {\bibfnamefont {J.}~\bibnamefont {Hare}}, \bibinfo {author}
  {\bibfnamefont {J.~M.}\ \bibnamefont {Raimond}}, \bibinfo {author}
  {\bibfnamefont {S.}~\bibnamefont {Haroche}}, \ and\ \bibinfo {author}
  {\bibfnamefont {W.}~\bibnamefont {Gawlik}},\ }\href@noop {} {\bibfield
  {journal} {\bibinfo  {journal} {Phys. Rev. A}\ }\textbf {\bibinfo {volume}
  {48}},\ \bibinfo {pages} {3991} (\bibinfo {year} {1993})}\BibitemShut
  {NoStop}%
\bibitem [{\citenamefont {Gray}\ \emph {et~al.}(1988)\citenamefont {Gray},
  \citenamefont {Zheng}, \citenamefont {Smith},\ and\ \citenamefont
  {Dunning}}]{gray:1988}%
  \BibitemOpen
  \bibfield  {author} {\bibinfo {author} {\bibfnamefont {D.~F.}\ \bibnamefont
  {Gray}}, \bibinfo {author} {\bibfnamefont {Z.}~\bibnamefont {Zheng}},
  \bibinfo {author} {\bibfnamefont {K.~A.}\ \bibnamefont {Smith}}, \ and\
  \bibinfo {author} {\bibfnamefont {F.~B.}\ \bibnamefont {Dunning}},\
  }\href@noop {} {\bibfield  {journal} {\bibinfo  {journal} {Phys. Rev. A}\
  }\textbf {\bibinfo {volume} {38}},\ \bibinfo {pages} {1601} (\bibinfo {year}
  {1988})}\BibitemShut {NoStop}%
\bibitem [{\citenamefont {Zimmerman}\ \emph {et~al.}(1979)\citenamefont
  {Zimmerman}, \citenamefont {Littman}, \citenamefont {Kash},\ and\
  \citenamefont {Kleppner}}]{zimmerman:1979}%
  \BibitemOpen
  \bibfield  {author} {\bibinfo {author} {\bibfnamefont {M.~L.}\ \bibnamefont
  {Zimmerman}}, \bibinfo {author} {\bibfnamefont {M.~G.}\ \bibnamefont
  {Littman}}, \bibinfo {author} {\bibfnamefont {M.~M.}\ \bibnamefont {Kash}}, \
  and\ \bibinfo {author} {\bibfnamefont {D.}~\bibnamefont {Kleppner}},\
  }\href@noop {} {\bibfield  {journal} {\bibinfo  {journal} {Phys. Rev. A}\
  }\textbf {\bibinfo {volume} {20}},\ \bibinfo {pages} {2251} (\bibinfo {year}
  {1979})}\BibitemShut {NoStop}%
\bibitem [{\citenamefont {Han}\ \emph {et~al.}(2006)\citenamefont {Han},
  \citenamefont {Jamil}, \citenamefont {Norum}, \citenamefont {Tanner},\ and\
  \citenamefont {Gallagher}}]{han:2006}%
  \BibitemOpen
  \bibfield  {author} {\bibinfo {author} {\bibfnamefont {J.}~\bibnamefont
  {Han}}, \bibinfo {author} {\bibfnamefont {Y.}~\bibnamefont {Jamil}}, \bibinfo
  {author} {\bibfnamefont {D.~V.~L.}\ \bibnamefont {Norum}}, \bibinfo {author}
  {\bibfnamefont {P.~J.}\ \bibnamefont {Tanner}}, \ and\ \bibinfo {author}
  {\bibfnamefont {T.~F.}\ \bibnamefont {Gallagher}},\ }\href@noop {} {\bibfield
   {journal} {\bibinfo  {journal} {Phys. Rev. A}\ }\textbf {\bibinfo {volume}
  {74}},\ \bibinfo {eid} {054502} (\bibinfo {year} {2006})}\BibitemShut
  {NoStop}%
\bibitem [{\citenamefont {Lindgard}\ and\ \citenamefont
  {Nielsen}(1977)}]{lindgard:1977}%
  \BibitemOpen
  \bibfield  {author} {\bibinfo {author} {\bibfnamefont {A.}~\bibnamefont
  {Lindgard}}\ and\ \bibinfo {author} {\bibfnamefont {S.~E.}\ \bibnamefont
  {Nielsen}},\ }\href@noop {} {\bibfield  {journal} {\bibinfo  {journal}
  {Atomic Data and Nuclear Data Tables}\ }\textbf {\bibinfo {volume} {19}},\
  \bibinfo {pages} {533} (\bibinfo {year} {1977})}\BibitemShut {NoStop}%
\bibitem [{\citenamefont {Hinds}(1991)}]{hinds:1991}%
  \BibitemOpen
  \bibfield  {author} {\bibinfo {author} {\bibfnamefont {E.~A.}\ \bibnamefont
  {Hinds}},\ }\href@noop {} {\bibfield  {journal} {\bibinfo  {journal} {Adv.
  Atom. Mol. Phys.}\ }\textbf {\bibinfo {volume} {28}},\ \bibinfo {pages} {237}
  (\bibinfo {year} {1991})}\BibitemShut {NoStop}%
\end{thebibliography}%

\end{document}